\pdfoutput=1
\documentclass[11pt]{article}

\usepackage[margin=1in]{geometry}
\usepackage{amsmath,amssymb}
\usepackage{graphicx}
\usepackage{booktabs}
\usepackage{siunitx}
\usepackage[hidelinks]{hyperref}
\usepackage{caption}
\usepackage{placeins}
\usepackage{microtype}
\DeclareSIUnit\angstrom{\text{\normalfont\AA}}
\DeclareSIUnit\hartree{Ha}

\graphicspath{{figures/}}

\newcommand{\Tc}{\ensuremath{T_c}}
\newcommand{\alphaF}{\ensuremath{\alpha^2F(\omega)}}
\newcommand{\omlog}{\ensuremath{\omega_{\log}}}
\newcommand{\wtwo}{\ensuremath{\omega_2}}
\newcommand{\chem}[1]{\ensuremath{\mathrm{#1}}}

\title{\bfseries Inverse design of superconductors and semiconductors with differentiable foundation models}

\author{
  Crist\'obal M\'endez,$^{1}$ Justin Tahmassebpur,$^{1}$ and Tom\'as A.\ Arias$^{2}$\\[3pt]
  \small $^{1}$School of Applied and Engineering Physics, Cornell University, Ithaca, New York 14853\\
  \small $^{2}$Department of Physics, Cornell University, Ithaca, New York 14853
}
\date{}

\begin{document}
\maketitle

\begin{abstract}
Designing a crystal for a target property requires choosing its elements,
stoichiometry, and atomic structure together, yet gradients exist only for the
continuous structural variables. We introduce differentiable Equiformer fast
effective-atom theory (DEFEAT), which lets a pretrained EquiformerV3
foundation model accept continuous species weights and site occupancies
without retraining on fractional data, thereby enabling gradient-based materials optimization. Gradients from fine-tuned property models and a
machine-learned energy penalty then guide updates to composition, atom count, atomic
positions, and lattice vectors in a single optimization. In a
search for phonon-mediated superconductors, density-functional perturbation
theory on 45 structurally novel designs gives isotropic Allen--Dynes critical
temperatures above 10~K for 39; the highest is 24.0~K for ZrTiNb$_2$CN$_3$,
and H$_4$LiPd$_3$ (23.2~K) lies on the static DFT hull. In searches for
photovoltaic absorbers, SCAN calculations confirm ten novel crystal structures
with band gaps within 20\% of a 1.34~eV target on their respective raw or
empirically corrected scales, all within 70~meV/atom of the DFT hull, and show
that the systematic gap error of the training functional propagates into the
designs unless the objective corrects for it. Quasiharmonic
free energies from machine-learned potentials further show that
finite-temperature competition with solid phases can reorder candidates
relative to the static hull.
\end{abstract}

\section{Introduction}

Finding a crystal with a prescribed property means deciding which elements to
combine, in what proportions, and in which structure. Atomic positions and
lattice vectors are continuous variables, but elemental identities and the
number of occupied sites are discrete, so a property cannot be differentiated
with respect to the chemical choices that most strongly control it.
Generative models and high-throughput screening avoid this difficulty by
proposing candidates in large numbers
\cite{gnome,mattergen,cdvae,prakash2026guided}; the target property then
enters as a sampling condition or a downstream filter. These workflows have
produced millions of candidate structures, but evaluation of the property alone does not give information on what changes would improve the performance of the candidate structure. Gradient-based optimization would supply
that information, provided chemistry and structure could be treated as
continuous variables of a single differentiable objective.

Continuous chemical representations bridge part of this gap. Recently introduced effective-atom
theory allows for optimization of site-specific mixtures of elements within density-functional
theory (DFT) \cite{feat,heat}, and alchemical extensions of machine-learned
interatomic potentials (MLIPs) make energies differentiable with respect to
composition and vacancy transformations \cite{nam2025alchemical}. Neither,
however, supplies inexpensive gradients of a functional property such as a
superconducting critical temperature or a band gap: effective-atom DFT remains
tied to electronic-structure calculations, and alchemical MLIPs predict
energies and forces. Meanwhile, equivariant foundation models trained on large
energy--force datasets learn transferable representations of atomic
environments that can be fine-tuned for other properties
\cite{uma,mlips,equiformerv3,jmp}. Inverse design therefore requires a
property model that is accurate on discrete crystals, differentiable with
respect to chemical identity and site occupancy, and coupled to an energy term
that keeps the optimization near thermodynamic stability.

Here we introduce differentiable Equiformer fast effective-atom theory
(DEFEAT), which meets these requirements with existing pretrained weights.
The key observation is that EquiformerV3 first encodes chemical identity
through a learned element-embedding table, an operation linear in a one-hot
species vector. Replacing that vector with continuous species weights
interpolates between learned elemental embeddings, and scaling message passing
by a site occupancy removes a site from the graph continuously. A model
trained only on discrete crystals therefore accepts fractional chemistry
without fractional training data. Moreover, we fine-tune EquiformerV3 to predict
electron--phonon parameters and semiconductor band gaps and couple these
predictors to an MLIP energy penalty, so that a single backward pass updates
chemical identity, occupancy, coordinates, and lattice vectors with respect to
both the target property and energetic stability. Annealing sharpens the
fractional representation toward discrete elements and vacancies, after which
the structure is relaxed. Because this local optimization depends on its
starting point, a Bayesian optimizer selects initial compositions.

We test the framework on two objectives and evaluate the resulting designs
with first-principles calculations. For phonon-mediated superconductors,
density-functional perturbation theory (DFPT) gives Allen--Dynes estimates
above \SI{10}{K} for 39 of 45 structurally novel designs, and
finite-temperature free energies show that the highest-\Tc{} candidates and
the most thermodynamically favored ones differ. For photovoltaic absorbers, we
target a band gap on both the raw SCAN scale and an empirically corrected
scale; SCAN calculations confirm ten novel crystal structures within 20\% of the target
on their respective scales,
and the comparison between scales shows how the error of the training
functional enters the designs.

\section{Results}

\subsection{Transfer learning for property prediction}

Gradient-based design is only as reliable as the property model that supplies
the gradients, so we first ask whether a foundation model trained on energies
and forces can predict electron--phonon parameters and band gaps accurately.
Labeled data for these properties are far scarcer than energy--force data.
We therefore fine-tune an EquiformerV3 model pretrained on OMat24, MPtrj, and sAlex \cite{equiformerv3,omat24}, whose
representations of local bonding environments are learned from far more
structures than any property dataset contains. Because the fine-tuned
predictors remain differentiable, they also supply property gradients with
respect to chemistry and structure at the cost of a single backward pass.

\begin{figure}[!htbp]
\centering
\includegraphics[width=\linewidth]{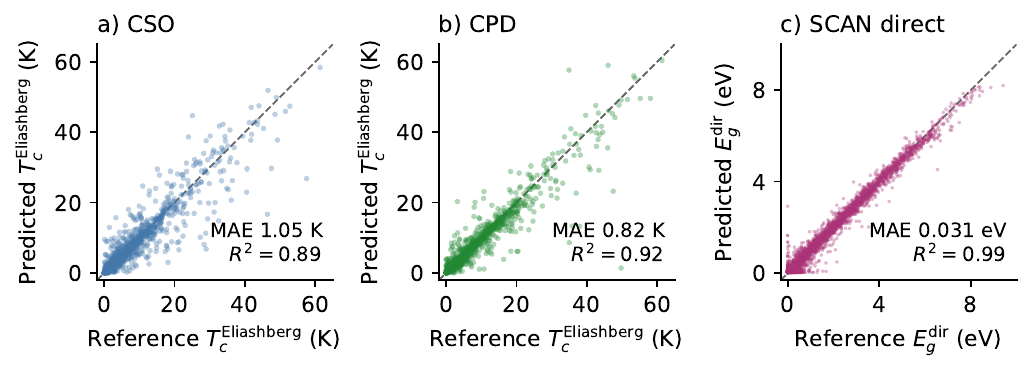}
\caption{Property prediction on held-out test sets. Predicted versus reference
values for (a) CSO \Tc{}, (b) CPD \Tc{}, and (c) the direct SCAN band gap.
The Eliashberg-based critical temperatures use the Allen--Dynes approximation,
with DFT electron--phonon parameters for the reference values and
directly predicted parameters for the model values.}
\label{fig:property-tests}
\end{figure}

For superconductivity, we follow the naming convention of Gibson et al.\
\cite{gibson}: crystal-structure-only (CSO) models use the atomic structure,
while coarse projected-phonon-DOS (CPD) models additionally use phonon
information. Rather than regressing \Tc{} directly, both predict the
electron--phonon coupling $\lambda$ and the frequency moments \omlog{} and
\wtwo{}, from which \Tc{} follows through the Allen--Dynes equation (Methods);
the predicted \Tc{} thus inherits the physical dependence of that equation on
coupling strength and phonon frequency scale. On held-out test structures, our
CSO model and the CPD ensemble reach \Tc{} mean absolute errors (MAEs) of
\SI{1.05}{K} and \SI{0.82}{K}, with $R^2=0.89$ and $0.92$, respectively
(Fig.~\ref{fig:property-tests}a,b). Gibson et al.\ report $R^2$ values of
$0.66$ and $0.79$ for the corresponding models; because the datasets and test
splits differ, this comparison is indicative rather than direct
(Supplementary Table~S1). Phonon information lowers the error further, but
CPD requires a phonon density of states at every step, and backpropagating
through its calculation is too slow to be practical inside the optimization.
CSO therefore supplies the differentiable search objective, and CPD reranks
the discrete candidates that pass it.

For band gaps, the choice of DFT functional affects both the fidelity and the
availability of training labels. We train PBE, SCAN, and
HSE06 models \cite{pbe,scan,hse06} on Matbench MP-gap, Alexandria-SCAN, and
SNUMAT, respectively \cite{matbench,alexandriascan,snumat}. PBE provides an
established benchmark: on fold~0 of Matbench MP-gap, our model achieves an MAE
of \SI{0.080}{eV}, compared with \SI{0.089}{eV} reported for JMP-L on the same
fold \cite{jmp}. HSE06 describes semiconductor gaps more accurately, but its
much smaller training set leads to larger prediction errors (Supplementary
Table~S2). SCAN provides the practical compromise for our search: an improved
semilocal description relative to PBE \cite{borlido2019}, a large training
set, and low prediction errors. Because photovoltaic absorption onset is set
by the direct gap, we use the SCAN direct-gap model, which reaches a test MAE
of \SI{0.031}{eV} and $R^2=0.99$ (Fig.~\ref{fig:property-tests}c).

\FloatBarrier

\subsection{Joint optimization of chemistry and structure}

\begin{figure}[!htbp]
\centering
\includegraphics[width=0.82\linewidth]{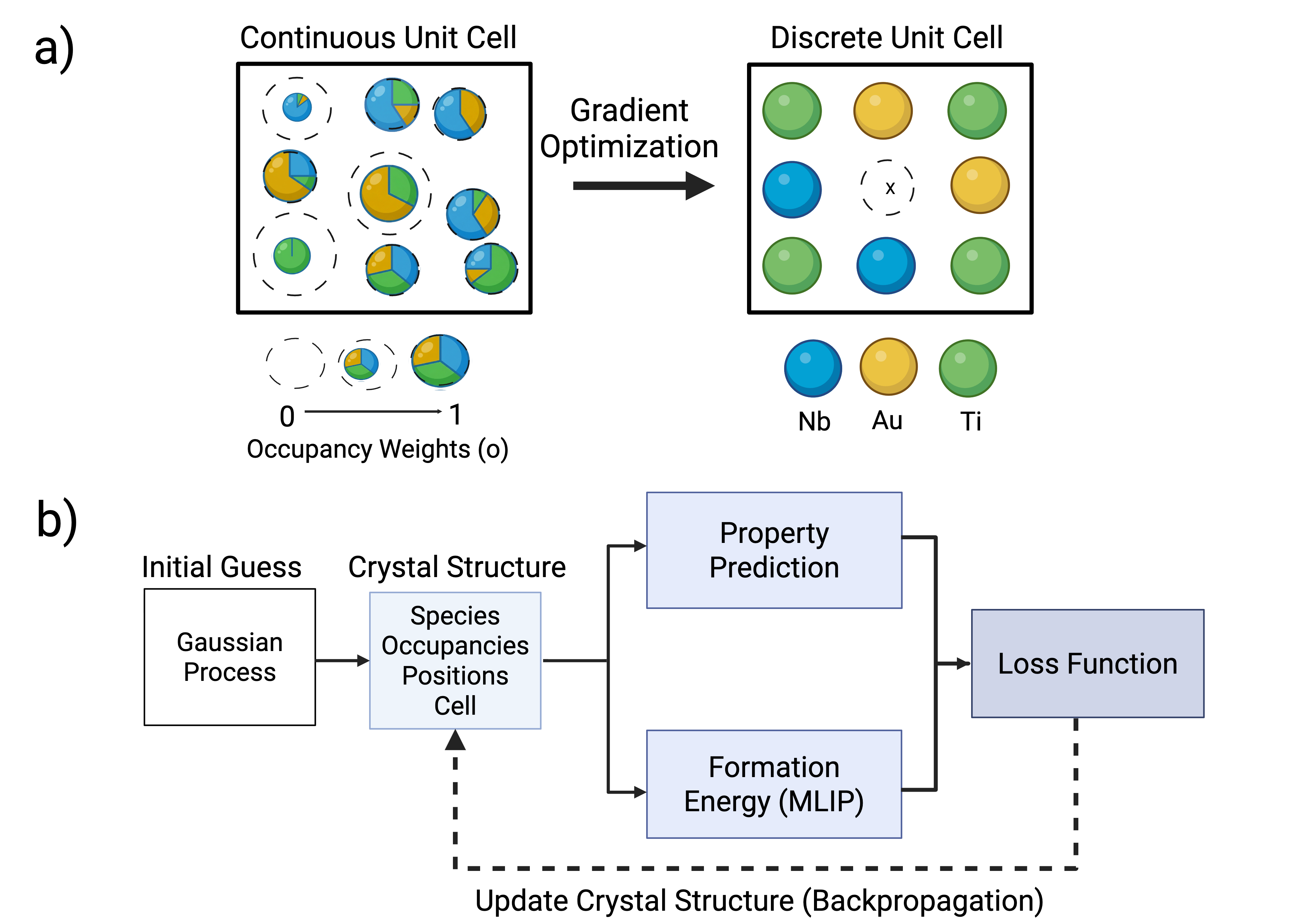}
\caption{Differentiable crystal design with DEFEAT.
(a) Each site carries species weights (pie fractions) and an occupancy $o$
(marker size). Optimization sharpens these variables toward discrete elemental
identities and occupied or vacant sites.
(b) A Gaussian-process model proposes starting compositions. Differentiable
property and energy models supply a combined objective whose gradients update
species weights, occupancies, atomic positions, and lattice vectors. The final
discrete structures are relaxed and evaluated.}
\label{fig:schematic}
\end{figure}

DEFEAT, which is based on effective-atom theory \cite{feat,heat}, represents each site $i$ by an occupancy $o_i$ and a normalized species
distribution $p_{iz}$, so that the weight of element $z$ at that site is
$W_{iz}=o_i p_{iz}$ (Fig.~\ref{fig:schematic}a). Changing $p_{iz}$ changes the
site's chemical identity; changing $o_i$ removes or activates sites within the
initialized cell. To evaluate these fractional states, the property models mix
their learned elemental embeddings according to $W$ and weight atomic
contributions by occupancy. Property gradients therefore act on composition
and atom count together with atomic positions and lattice vectors, so a single gradient-guided trajectory can change both stoichiometry and structure. During optimization,
the species weights and occupancies are progressively sharpened toward discrete
elements and occupied or vacant sites, and the final crystal is relaxed at fixed
elemental composition. The fractional states serve only as an optimization space:
all reported properties below are evaluated on the resulting discrete structures.

A property objective alone has no reason to remain near thermodynamic
stability. Stability is commonly assessed by screening after property-driven generation. To favor energetically viable candidates during the search, DEFEAT instead adds an energetic penalty to the property objective, favoring
structures with low energy relative to reference phases (Materials Project \cite{materialsproject} phases recomputed with the MLIP for
superconductors; Alexandria-SCAN formation energies for semiconductors, with a
consistent reevaluation after the search) throughout the
optimization (Fig.~\ref{fig:schematic}b; Methods). For superconductors, the property term maximizes the CSO-derived
\Tc{}; for semiconductors, it penalizes deviation from the target direct gap.
Because each trajectory is a local optimization of a non-convex objective, the starting composition for each
trajectory is proposed by Bayesian optimization \cite{bayesopt} to enable a more global search. To initialize the Bayesian optimization, the underlying
Gaussian-process models are seeded with filtered Materials Project structures whose properties and
energies were evaluated with the same property and energy models as the search. Each
final discrete search outcome is then used to update the Gaussian processes employed in the optimization (Methods).

\FloatBarrier

\subsection{Superconductor discovery and first-principles evaluation}

\begin{figure}[!htbp]
\centering
\includegraphics[width=0.92\linewidth]{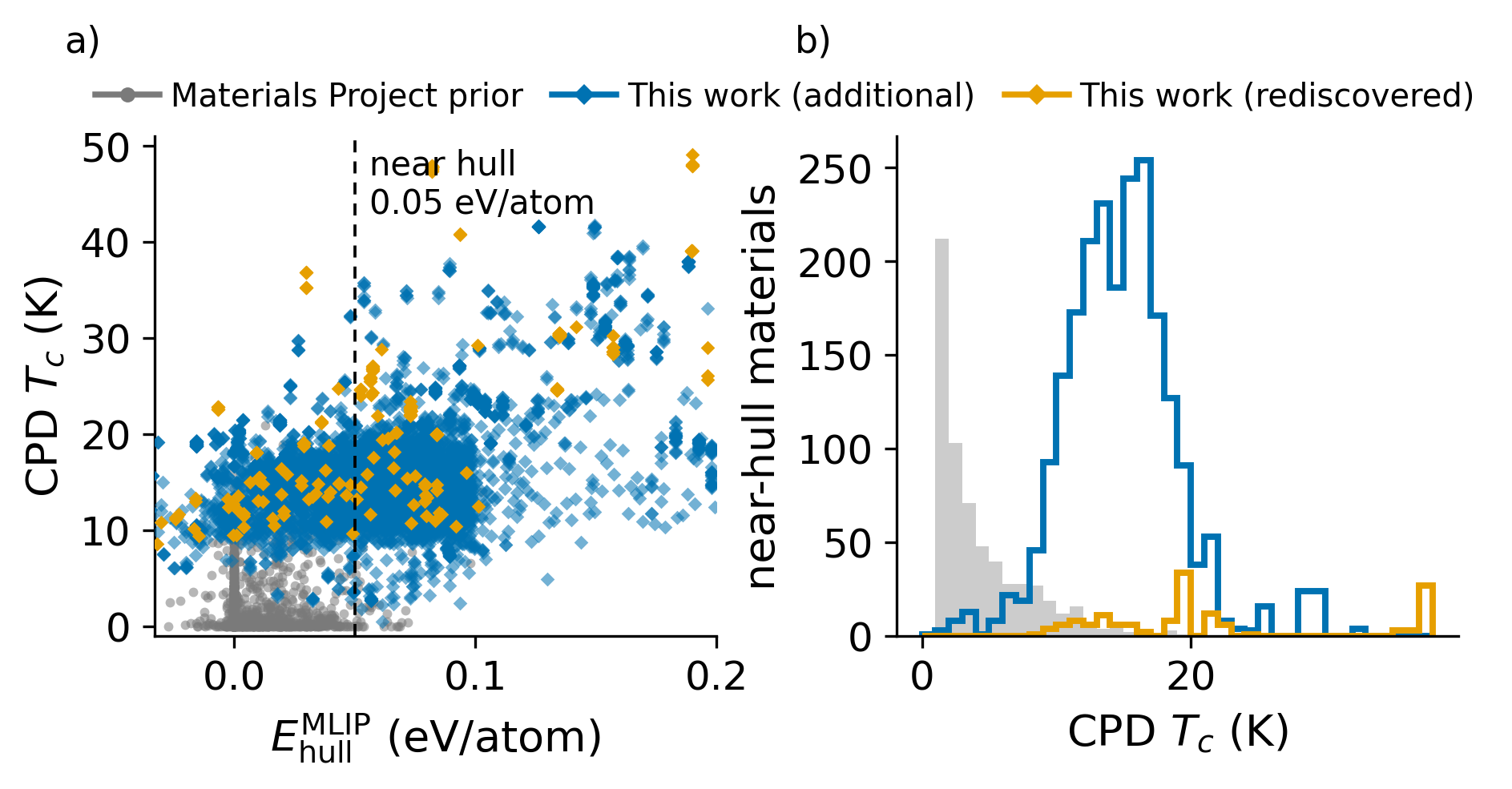}
\caption{Predicted superconducting properties across the search population.
(a) CPD \Tc{} versus the signed MLIP energy offset from the reference hull for
the Materials Project prior (gray) and generated candidates classified as
stable by the MLIP phonon screen. Blue and orange distinguish
candidates without and with a composition--space-group match in the
Alexandria electron--phonon dataset, respectively; this preliminary novelty
screen is distinct from the structure matching used for the validated designs.
The dashed line marks the \SI{0.05}{eV/atom} screening threshold.
(b) Counts of near-hull candidates as a function of predicted \Tc{}; the
0--1 K bin of the Materials Project prior is omitted for readability.}
\label{fig:tcdisc}
\end{figure}

We first ask whether property gradients lead to superconductors that survive
first-principles evaluation. The consolidated search generated 66{,}185
distinct structures spanning 18{,}068 reduced compositions. The final follow-up campaign
selected candidates with CSO-predicted $\Tc>\SI{9}{K}$ and MLIP hull distances of
at most \SI{0.10}{eV/atom}, screening those at or below
\SI{0.05}{eV/atom} first. Across all campaigns, 14{,}233 structures were
screened with CPD; 8{,}070 passed the MLIP dynamical-stability criteria (no significant imaginary phonon modes),
including 4{,}777 with
predicted $\Tc>\SI{10}{K}$ (Methods). Their distribution extends into regions
of elevated predicted \Tc{} beyond the filtered Materials Project prior
(Fig.~\ref{fig:tcdisc}), which supplies initialization and chemical
context.

The generated structures include transition-metal carbides, nitrides and
carbonitrides, hydride--nitrides, and alloys. The search also recovers
established superconductors such as NbN, ZrN, TiN, NbC, and TaC
\cite{nbn,zrn,tin,nbctac}, which serve as internal controls. Because such
compounds may also appear in the Materials Project prior, these recoveries
indicate consistency with known chemistry rather than independent discovery.
Structure matching against the reference
databases (Methods) separates them from new compositions and polymorphs.

Beyond rediscovering structures in the reference databases examined
here, we identified 45 additional candidates: 33 new compositions and
12 new polymorphs of compositions present in those databases. We evaluated these candidates using first-principles
electron--phonon calculations. Candidates were selected for high
predicted \Tc{} and low MLIP hull distance, with preference for
higher-symmetry structures, whose DFPT calculations converge more
reliably. Of these, 39 have Allen--Dynes $\Tc>\SI{10}{K}$ and three
exceed \SI{20}{K}. Table~\ref{tab:sc-cands} lists the leading
candidates, together with two additional designs used in the
finite-temperature analysis. The highest estimate is \SI{24.0}{K}
for \chem{ZrTiNb_2CN_3}, and three related multi-cation carbonitrides
also exceed \SI{15}{K}.

Static DFT reference-hull calculations provide an independent energetic check
for all 15 listed candidates; campaign-wide screening statistics use MLIP hull
estimates, whereas Table~\ref{tab:sc-cands} reports the DFT assessments. Six
candidates lie on the assessed hull, including \chem{H_4LiPd_3}
($\Tc=\SI{23.2}{K}$) and \chem{Mo_4CN}, and three more lie within
\SI{50}{meV/atom}. The two families separate energetically: four of the six
on-hull designs contain hydrogen, whereas all four multi-cation carbonitrides
in Table~\ref{tab:sc-cands} lie 56--\SI{87}{meV/atom} above the hull,
including the leading \chem{ZrTiNb_2CN_3} at \SI{79}{meV/atom}. In the
carbonitrides, the highest estimated \Tc{} and static stability therefore do
not coincide, a trade-off we examine at finite temperature in
Section~\ref{sec:qha}.

All critical temperatures reported here are estimates from a harmonic,
isotropic Allen--Dynes treatment with $\mu^*=0.10$, which omits anharmonicity
and gap anisotropy. The anisotropic two-gap superconductivity of
\chem{MgB_2}, the highest-\Tc{} conventional superconductor at ambient
pressure, lies outside this treatment \cite{choi}, so its experimental
\SI{39}{K} \cite{mgb2} is not a direct benchmark for the estimates here.

\begin{table}[!htbp]
\centering
\caption{Leading superconducting designs. Critical temperatures are Allen--Dynes
estimates from DFPT with $\mu^*=0.10$. DFT hull distances use the static PBEsol
reference assessment described in Methods. The dagger ($\dagger$) marks a new
polymorph of a composition present in the reference databases; all other
entries are new compositions.}
\label{tab:sc-cands}
\small
\begin{tabular}{lccc}
\toprule
Material & Sp.\ gp. & $T_c^{\mathrm{DFPT/AD}}$ (K) &
$E_{\mathrm{hull}}^{\mathrm{DFT}}$ (meV/atom) \\
\midrule
\chem{ZrTiNb_2CN_3} & 123 & 24.0 & 79 \\
\chem{H_4LiPd_3} & 6 & 23.2 & 0 \\
\chem{NbH_2}$^{\dagger}$ & 129 & 22.0 & 86 \\
\chem{Mo_2H_3} & 115 & 17.4 & 0 \\
\chem{Mo_5CN_2} & 12 & 16.9 & 48 \\
\chem{ZrTi_2NbCN_3} & 123 & 16.7 & 87 \\
\chem{TcMo_2H_3} & 71 & 16.7 & 0 \\
\chem{TiNb_3C_2N_2} & 123 & 15.8 & 66 \\
\chem{TiNb_3C_3N} & 107 & 15.5 & 56 \\
\chem{V_2H_2N} & 164 & 15.3 & 51 \\
\chem{H_3Tc_4} & 139 & 15.2 & 0 \\
\chem{LiTiNb_2N_4} & 123 & 15.1 & 0 \\
\chem{Tc_7N} & 148 & 15.1 & 18 \\
\chem{Mo_4CN} & 10 & 14.7 & 0 \\
\chem{Zr_3TiN_4} & 65 & 11.2 & 43 \\
\bottomrule
\end{tabular}
\end{table}

\FloatBarrier
\subsection{Temperature-dependent competition with solid phases}
\label{sec:qha}

Computational discovery workflows typically assess stability through the
static, zero-temperature convex hull \cite{gnome,mattergen}. Synthesis,
however, takes place at elevated temperature, where vibrational free energy
can reorder competing phases: a candidate above the static hull may become
favored, and one on it may not remain so. A finite-temperature assessment
requires phonon calculations for the candidate and for every competing phase
across a range of volumes, a cost that has so far kept it out of the discovery and filtering stages.
Machine-learned interatomic potentials reduce the cost of vibrational free energy calculations sufficiently to make such filtering
practical.

Of the 45 designs evaluated with DFPT, ten admit a usable finite-temperature comparison within the quasiharmonic approximation (QHA), which includes zero-point motion, vibrational free energy, and volume relaxation \cite{uma,phonopy,togo2023}. For these ten designs, the candidate and all phases in its controlling competing mixtures remain dynamically stable within the MLIP over a sufficiently broad volume range for quasiharmonic fitting (Methods). For the remaining 35 designs, the candidate or a controlling competitor fails these stability or fitting criteria, preventing a reliable quasiharmonic free-energy evaluation. For the ten eligible designs, we evaluated free energies over their computationally accessible temperature ranges, up to a maximum of \SI{1200}{K}, comparing against a reference pool of 129 Materials Project phases and additional evaluated structures (Methods). At each temperature, we compute $\Delta G$, the candidate free energy minus that of the lowest-free-energy, composition-conserving mixture of available competing solids (Methods).

\begin{figure}[!htbp]
\centering
\includegraphics[width=0.84\linewidth]{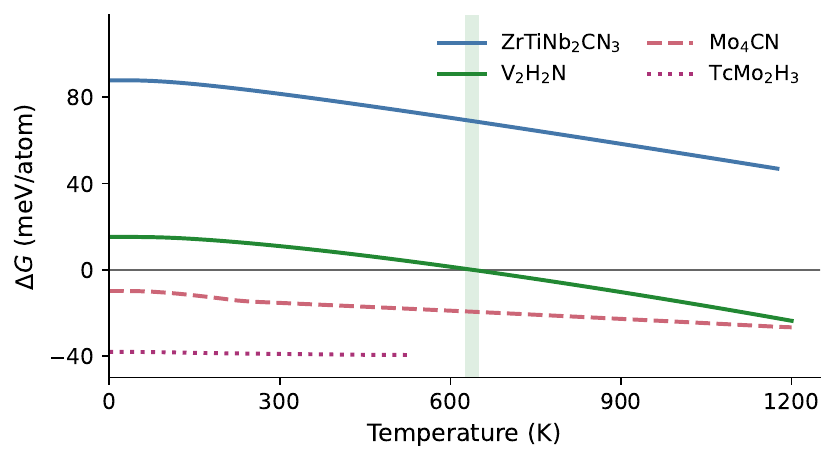}
\caption{Temperature-dependent competition for selected superconducting designs.
Signed MLIP-QHA free-energy differences are measured relative to the
lowest-free-energy mixture of available competing solids at each temperature,
conserving all elements. Negative values favor the candidate within this solid
pool. Values at \SI{0}{K} differ from Table~\ref{tab:sc-cands} because they
use UMA in place of PBEsol, include zero-point motion and volume relaxation,
use a different reference pool, and are not clipped at zero. Curves are restricted to usable temperatures with coverage of the identified
controlling phases; the shorter \chem{TcMo_2H_3} curve reflects this restriction.
The shaded interval marks the sampled sign change of \chem{V_2H_2N}.}
\label{fig:sc-stability}
\end{figure}

Figure~\ref{fig:sc-stability} shows the three eligible candidates with the lowest free-energy differences relative to their competing phases, together with \chem{ZrTiNb_2CN_3}, which has the highest estimated \Tc{}; Supplementary Fig.~S1 and Table~S4 summarize all ten over their accessible quasiharmonic ranges, including their competing mixtures. We find that \chem{Mo_4CN} and \chem{TcMo_2H_3}, both on the static DFT hull, are favored over their competing solid phases already at zero temperature and remain so throughout their accessible MLIP-QHA temperature ranges. \chem{V_2H_2N}, which lies \SI{51}{meV/atom} above the static DFT hull and would be deprioritized by a conventional screen, has an MLIP-QHA free-energy difference that changes sign between \SI{625}{K} and \SI{650}{K}, becoming favored over its competing mixture, \chem{VN} + \chem{VH_2}, above this bracket (Supplementary Table~S4). This result suggests that synthesis at elevated temperature followed by rapid quenching may provide a route to this phase, provided decomposition can be kinetically suppressed during cooling. By contrast, our highest-$T_c$ candidate, \chem{ZrTiNb_2CN_3}, remains above its competing phases throughout the accessible temperature range, although its relative stability improves with increasing temperature. Given uncertainties in the zero-temperature energies, synthesis at elevated temperature followed by quenching remains a possibility worth exploring, but is not established by these calculations. Future work could also examine whether high-pressure gas environments shift the relative stability of these candidates, particularly through the hydrogen and nitrogen chemical potentials imposed by \chem{H_2} and \chem{N_2}.

\FloatBarrier
\subsection{Targeting a photovoltaic band gap}

To test DEFEAT beyond superconductivity, we consider a different design objective in which, instead of maximizing a property, we search for materials with a property matching a prescribed value. We also consider how to correct for known discrepancies between the training databases and experiment. In particular, we use DEFEAT to search for semiconductors with band gaps near \SI{1.34}{eV}, approximately the Shockley--Queisser optimum for single-junction photovoltaics \cite{shockley,ruhle}. Because semilocal DFT systematically underestimates band gaps, directly targeting this value on the raw computational scale does not target the same experimental gap. We therefore ran searches on both the raw SCAN direct-gap scale and an empirically corrected scale,
\begin{equation}
E_g^{\mathrm{corr}}=1.43E_g^{\mathrm{SCAN,dir}}-\SI{0.071}{eV},
\label{eq:gap-correction}
\end{equation}
obtained by inverting the benchmark regression of Borlido et al.\
\cite{borlido2019}, $E_g^{\mathrm{SCAN}}=0.70E_g^{\mathrm{exp}}+\SI{0.05}{eV}$. This correction accounts for the regression's slope of 0.70 and its small offset, which together imply an underestimation that grows with the experimental gap. On the corrected scale, the target corresponds to a SCAN gap of approximately \SI{0.99}{eV}. The correction is an empirical, population-level mapping and does not account for material-specific deviations.

We ran DEFEAT independently for the raw and corrected band-gap
targets. Among all candidates, 45 raw-target and nine corrected-target compositions lay
within \SI{50}{meV/atom} of the UMA hull and within 20\% of the target gap on
the corresponding scale. \emph{Ab initio} SCAN calculations on the leading 20 raw-target and
all nine corrected-target candidates then confirmed gaps within 20\% of the target for 15 and six candidates,
respectively. From these, five candidates for each objective were selected by hull distance
and target proximity for further evaluation by SCAN relaxation and DFT hull assessment
(Table~\ref{tab:bg-cands}). All ten remain within 20\% of the target on the
scale they were optimized for and within \SI{70}{meV/atom} of the DFT
reference hull. None match structures in the Materials Project,
Alexandria-SCAN, or SNUMAT; seven are new compositions and three are new
polymorphs. \chem{NaSbTeSe} matches the raw target within \SI{0.002}{eV}, while
\chem{Na_2Se_2S}, \chem{K_5AgAs_2}, and \chem{BaSnS_2} lie within
\SI{10}{meV/atom} of the hull.

Comparing the two scales shows why the choice of target matters. Candidates
from the raw search, with SCAN gaps between 1.09 and \SI{1.57}{eV}, have
corrected gaps of 1.49 to \SI{2.17}{eV}, above the photovoltaic optimum.
The corrected search instead yields SCAN gaps between 0.89 and \SI{1.17}{eV},
which place the corrected gaps within the target window. Because the optimization
follows whichever scale it is given, the systematic error of the underlying
functional propagates into the designs unless it is accounted for in the
objective.

\begin{table}[!htbp]
\centering
\caption{Representative semiconductor designs from both target scales.
Direct gaps are evaluated on SCAN-relaxed structures; corrected gaps use
Eq.~\eqref{eq:gap-correction}, applied to unrounded SCAN gaps. Both gap
columns are displayed to three decimal places; target-window membership is
evaluated before rounding. Bold values indicate the scale targeted in each
search. Daggers ($\dagger$) mark new polymorphs of compositions present in
the reference databases; all other entries are new compositions.}
\label{tab:bg-cands}
\small
\begin{tabular}{lccc}
\toprule
Material & $E_g^{\mathrm{SCAN,dir}}$ (eV) & $E_g^{\mathrm{corr}}$ (eV) &
$E_{\mathrm{hull}}^{\mathrm{DFT}}$ (meV/atom) \\
\midrule
\multicolumn{4}{l}{\textit{Raw-target}} \\
\chem{Na_2Se_2S} & \textbf{1.569} & 2.173 & 1 \\
\chem{BaSnS_2}$^{\dagger}$ & \textbf{1.484} & 2.051 & 10 \\
\chem{NaSe_2}$^{\dagger}$ & \textbf{1.089} & 1.486 & 12 \\
\chem{KNaSe_4} & \textbf{1.276} & 1.754 & 15 \\
\chem{NaSbTeSe} & \textbf{1.342} & 1.848 & 17 \\
\multicolumn{4}{l}{\textit{Corrected-target}} \\
\chem{K_5AgAs_2} & 1.164 & \textbf{1.594} & 2 \\
\chem{Na_5TeAs} & 0.892 & \textbf{1.204} & 18 \\
\chem{Na_5AgP_2} & 1.086 & \textbf{1.482} & 20 \\
\chem{K_3Sn_3As} & 1.173 & \textbf{1.607} & 54 \\
\chem{KSe_2}$^{\dagger}$ & 1.081 & \textbf{1.474} & 68 \\
\bottomrule
\end{tabular}
\end{table}

\FloatBarrier
\section{Discussion}

The central result of this work is that gradients from a fine-tuned foundation model, extended to fractional chemistry, can drive the joint optimization of composition, occupancy, and structure toward candidates that hold up under first-principles evaluation. In the superconducting search, 39 of 45 designs selected for DFPT have Allen--Dynes estimates above \SI{10}{K}, and the most productive regions were multi-cation transition-metal carbonitrides and hydrogen-containing compounds. In the band-gap search, the same optimization matched a prescribed value rather than maximizing one, and the comparison between target scales showed that the optimization reproduces the systematic error of its training functional: a design is only as accurate as the scale on which the objective is defined. However, we also demonstrated how a known empirical relationship between computed and experimental values can be incorporated into the objective to guide the search toward the desired experimental target.

To help identify synthesizable candidates, we introduced an energetic penalty that steers the search toward the hull, although a trade-off between property optimization and stability remains. Indeed, the highest-\Tc{} carbonitrides lie 56--\SI{87}{meV/atom} above the static DFT hull, but we also identify six compounds on the assessed DFT hull, four of which contain hydrogen. Finite-temperature effects further alter these rankings: \chem{V_2H_2N}, which a static screen would deprioritize, becomes favored over its competing solid phases above roughly \SI{650}{K} and may be recoverable by quenching. Although \chem{ZrTiNb_2CN_3} does not become favored within its accessible quasiharmonic temperature range, its free-energy excess relative to its competing solid phases starts at approximately \SI{0.09}{eV/atom} and decreases with increasing temperature. Static hull distance is therefore an incomplete guide to which designs merit synthesis. To address this limitation, we have shown that MLIP-based quasiharmonic comparisons make finite-temperature screening computationally affordable.

DEFEAT differs from property-conditioned diffusion models \cite{mattergen,prakash2026guided} in how it reaches candidates. Diffusion models sample from a learned structural distribution steered toward a target, whereas DEFEAT starts from a proposed composition and follows explicit property and energy gradients with respect to species, occupancy, and geometry, providing a more interpretable optimization trajectory. The approaches can be combined: diffusion-generated structures could serve as starting points for gradient refinement. A quantitative comparison of the success rates or computational efficiency of DEFEAT and diffusion methods would require shared evaluation criteria and matched computational budgets, which the present study does not provide.

Beyond superconductors and semiconductors, DEFEAT offers a route to designing materials for new target properties and combinations of properties. Its differentiable property predictors and compatible energy model can build on existing pretrained weights, allowing new targets to be introduced through fine-tuning without changing the underlying optimization. Multiple property objectives and constraints can then guide the same search, opening the possibility of jointly optimizing electronic, optical, mechanical, or transport behavior while accounting for energetic stability and restrictions on elemental composition. Such searches could reveal new chemistries and structures that bring previously unexplored combinations of material functionality within reach.

\section{Methods}

\subsection{Effective-atom property models}

All property models are built within the FairChem framework \cite{fairchem}
and use the EquiformerV3 equivariant graph network \cite{equiformerv3},
initialized from the public \texttt{omat24-mptrj-salex} checkpoint
\cite{omat24}.
UMA-s-1p2 with its inorganic-materials task \cite{uma} supplies candidate
energies in the search objective, the superconductor reference hull (Materials
Project phases recomputed before optimization), post-search UMA hull
assessments, and finite-temperature energies; the semiconductor search instead
uses Alexandria-SCAN reference energies during optimization
(Section~\ref{sec:search}); phonon densities of states for screening
are computed with phonon-fine-tuned Nequix \cite{nequix,pft}.

We modified FairChem and EquiformerV3 to accept a weight matrix
$W\in\mathbb{R}^{N\times Z}$ for $N$ sites and $Z$ allowed elements in place
of one-hot species. Node embeddings are obtained by multiplying $W$ by the
learned element-embedding table; edge features and message passing are
weighted by the site occupancies $o_i=\sum_z W_{iz}$; sites with $o_i=0$ are
masked; and the graph-level readout uses an occupancy-weighted mean.
One-hot, unit-occupancy rows recover the conventional input, so the extended
model reproduces the trained model exactly on discrete crystals, and a
fractional row interpolates linearly between learned elemental embeddings. Species and
occupancy gradients are obtained by automatic differentiation of the model
output with respect to $W$.

The superconducting models predict $\lambda$, \omlog{}, and \wtwo{}, defined as moments of
the Eliashberg spectral function,
\begin{equation}
\begin{aligned}
\lambda &= 2\!\int_0^\infty \frac{\alphaF{}}{\omega}\,d\omega, \quad
\omlog{} = \exp\!\Big[\tfrac{2}{\lambda}\!\int_0^\infty
   \tfrac{\alphaF{}}{\omega}\ln\omega\,d\omega\Big], \\
\wtwo{} &= \Big[\tfrac{2}{\lambda}\!\int_0^\infty
   \alphaF{}\,\omega\,d\omega\Big]^{1/2},
\end{aligned}
\end{equation}
and \Tc{} follows from the Allen--Dynes equation \cite{allendynes,mcmillan},
\begin{equation}
\Tc{} =
\frac{f_1 f_2\,\omlog{}}{1.2}
\exp\!\Big[
\frac{-1.04(1+\lambda)}
{\lambda-\mu^*(1+0.62\lambda)}
\Big],
\end{equation}
with the strong-coupling and spectral-shape factors $f_1$ and $f_2$ defined
by Allen and Dynes \cite{allendynes}, $\mu^*=0.10$, and \omlog{} in
kelvin. \Tc{} is proportional to the characteristic phonon frequency \omlog{}
and depends exponentially on $\lambda$ through the competition between
phonon-mediated attraction and the screened Coulomb repulsion $\mu^*$, so
small errors in $\lambda$ at weak coupling produce large relative errors in
\Tc{}. The spectral-shape factor $f_2$ depends on the ratio
$\omega_2/\omega_{\log}$, motivating prediction of the second frequency
moment as well as \omlog{} and $\lambda$. The same definitions are used in the
first-principles evaluation. Training data come from the Alexandria
electron--phonon dataset \cite{alexandriaeph}; after removing dynamically
unstable compounds and entries with incomplete spectral data, 21{,}988
structures remain, split 80/10/10 into training, validation, and test sets.
The CPD model encodes the global projected phonon density of states and 14
derived phonon features, fused with the crystal embedding through a residual
multilayer perceptron. Its phonon inputs come from DFT for training and
testing and from Nequix for screening, and its predictions are averaged over
seven independently seeded models.

All models are trained with AdamW. CSO and each CPD member are initialized
directly from the pretrained checkpoint, with the backbone and property heads
trained jointly from the first stage. Both use three stages with head learning
rates of $3\times10^{-5}$, $5\times10^{-6}$, and $10^{-6}$. CSO uses these
same rates for the backbone; CPD uses backbone rates of $10^{-5}$,
$2\times10^{-6}$, and $5\times10^{-7}$. The stage epoch limits are 40, 40,
and 20 with early stopping, and the effective batch size is 8. Band-gap
models use a batch size of 16. The final-stage backbone and head learning
rates are $10^{-6}$ and $5\times10^{-5}$ for PBE, $8\times10^{-7}$ and
$4\times10^{-5}$ for SCAN, and $2\times10^{-7}$ and $8\times10^{-6}$ for HSE06.

\subsection{DEFEAT optimization and initialization}

Each trajectory initializes eight candidate sites on a periodic
$2\times2\times2$ cubic grid. Each site carries an occupancy logit $u_i$ and
species logits $\mathbf{v}_i$, mapped to
$o_i=\sigma(u_i/T_{\mathrm{occ}})$ and
$p_{iz}=\operatorname{softmax}_z(v_{iz}/T_{\mathrm{comp}})$; separate
variables control site displacements and lattice vectors. The loss is
$\mathcal{L}=\mathcal{L}_{\mathrm{property}}+\mathcal{L}_{\mathrm{hull}}
+\mathcal{L}_{\mathrm{reg}}$. The superconducting property term is
$-0.20\,\Tc{}/(\SI{10}{K})$, using CSO-predicted \Tc{}; the semiconductor
term is a Gaussian-window penalty of weight 5.0 and width \SI{0.30}{eV}
centered at the target gap. The hull term is a softplus barrier on the
composition-weighted energy above the reference hull, with a target of
\SI{0.035}{eV/atom} and a softplus smoothing scale of \SI{0.025}{eV/atom}.
Regularization uses a quadratic penalty on deviations of the fractional
species counts $n_z=\sum_i W_{iz}$ from the proposed counts $n_z^\star$,
$w_{\mathrm{sp}}\sum_z(n_z-n_z^\star)^2$, in composition-constrained searches.
Unconstrained campaigns instead use lower-count penalties for selected species
or element groups and a softplus penalty on a deficit in the compositional
entropy $-\sum_z x_z\ln x_z$, where $x_z=n_z/\sum_{z'}n_{z'}$.
An occupancy-weighted short-distance penalty discourages overlapping sites.
The campaign configurations specify the active terms and their weights.

Variables are updated with Adam at learning rates of 0.04 for positions and
lattice parameters and 0.03 for occupancy and species logits. Each trajectory
explores at $T_{\mathrm{occ}}=T_{\mathrm{comp}}=1.5$, then anneals through
four constant-temperature stages at 0.429, 0.122, 0.035, and 0.01. Sites are
then mapped to discrete elements or vacancies, and the structure is relaxed
at fixed composition.

Starting compositions are proposed by Bayesian optimization (BO). Compositions are
represented by elemental fractions, species-presence indicators, and
normalized cell size, and Gaussian processes with radial-basis-function kernels, initially
conditioned on the Materials Project prior, model the target property and the
MLIP hull energy. For superconductors, the acquisition function combines an
upper confidence bound on predicted \Tc{}, weighted by the probability of
lying below the hull threshold, with a novelty term that decays over
iterations and favors compositions far from those already evaluated; for
semiconductors, it maximizes the expected proximity to the target gap. Each
selected composition is optimized from several random restarts, with fixed
step budgets for exploration, annealing, and final relaxation, and early
stopping.

\subsection{Search spaces and screening}
\label{sec:search}

The superconductor search used 25 elements: H, Li, B, C, N, Mg, Al, Sc, Ti, V, Ni, Cu, Zn,
Y, Zr, Nb, Mo, Tc, Ru, Rh, Pd, La, Ta, Ir, and Pt. Each appears in more than
1{,}000 structures of the Alexandria electron--phonon dataset, ensuring
training coverage across the search space. The BO prior comprises the 1{,}886
Materials Project (MP) entries composed only of these elements that are metallic, contain at
most eight atoms and four species per cell, and lie within
\SI{0.05}{eV/atom} of the Materials Project hull. CPD phonon inputs were
computed with Nequix and Phonopy \cite{phonopy} using supercells with a
minimum length of \SI{8}{\angstrom}, a $12^3$ mesh, and 512 bins.
Candidates are classified as dynamically stable when the minimum frequency is
at least \SI{-0.10}{THz} and the imaginary-mode fraction at most 0.001.

The semiconductor search used 16 elements: O, S, Se, Te, P, As, Sb, Cu, Ag, Na, K, Ba, In,
Sn, Bi, and I. The BO prior contains Materials Project structures with
model-predicted direct gaps near the target, at most eight atoms and four
species per cell, and an MP hull distance below \SI{0.05}{eV/atom}. During
the search, the hull term combined UMA candidate formation energies with
Alexandria-SCAN reference formation energies; because this mixed proxy is not
a consistent hull distance, all candidates and competing phases were
reevaluated with UMA under a common energy convention before selection.
Candidates within \SI{0.05}{eV/atom} of the UMA hull and with predicted gaps
within $\pm20\%$ of the target on the corresponding scale were retained,
excluding molecular structures and database matches. The leading candidates
were screened with SCAN, and those retaining target gaps were selected by
hull distance and target proximity for SCAN relaxation and DFT hull
assessment.

\subsection{Novelty assessment}

Superconducting candidates were compared with the Materials Project and the
Alexandria electron--phonon dataset; semiconductor candidates with the
Materials Project, Alexandria-SCAN, and SNUMAT. Element-preserving structure
matching \cite{pymatgen} of the relaxed candidates included primitive cells, volume scaling,
and supercells, with length, site, and angle tolerances of 0.2, 0.3, and
$5^{\circ}$, and a broader 0.3, 0.5, and $10^{\circ}$ check.
Unmatched structures are classified as new compositions when absent from
these datasets and as new polymorphs otherwise.

\subsection{First-principles evaluation}

Electron--phonon calculations used Quantum ESPRESSO \cite{qe} with the
PBEsol functional \cite{pbesol}, pseudopotentials, smearing, and $k$/$q$-sampling protocol
of the Alexandria electron--phonon labels. Band gaps were computed with SCAN
in JDFTx \cite{jdftx} on SCAN-relaxed structures.

\subsection{DFT hull assessment}

Tables~\ref{tab:sc-cands} and \ref{tab:bg-cands} report nonnegative hull
distances after including the candidate, whereas Fig.~\ref{fig:tcdisc} shows
signed MLIP offsets, which are negative below the reference hull.
Superconducting checks use PBEsol in Quantum ESPRESSO with PseudoDojo \cite{pseudodojo}
scalar-relativistic v0.4.1 stringent pseudopotentials, cutoffs of 140 and
560 Ry, Methfessel--Paxton smearing of 0.02 Ry, and a $k$-point spacing of
\SI{0.20}{\per\angstrom}, against the MP-stable phases in each chemical
system. Semiconductor checks use SCAN in JDFTx with SG15 ONCV PBE-1.0
pseudopotentials \cite{sg15}, a \SI{30}{\hartree} cutoff, cold smearing of
\SI{0.001}{\hartree}, and a $k$-point spacing of \SI{0.35}{\per\angstrom}.
Candidates are relaxed with electronic and geometry energy thresholds of
\SI{1e-5}{\hartree} and \SI{5e-5}{\hartree}, and final static energies use
\SI{1e-6}{\hartree}; all energies are nonmagnetic and scalar-relativistic.
Reference phases, at fixed geometries, are MP-stable structures and
UMA-predicted decomposition products.

\subsection{Finite-temperature free energies}

The reference pool contains 129 Materials Project phases, supplemented by a
MoN structural variant and other evaluated designs in each chemical system.
For each phase, UMA-s-1p2 static energies at seven volume factors from 0.94
to 1.06 (BFGS relaxation to \SI{0.005}{eV/\angstrom}) were combined with
finite-displacement phonons (minimum supercell height \SI{9}{\angstrom},
\SI{0.01}{\angstrom} displacements, $20^3$ mesh) and Phonopy vibrational free
energies \cite{phonopy,togo2023}, with denser sampling for selected phases.
At $p=\SI{1}{atm}$,
\begin{equation}
G_i(T,p)=\min_V\left[E_i^{\mathrm{UMA}}(V)+F_i^{\mathrm{vib}}(T,V)+pV\right],
\end{equation}
evaluated from 0 to \SI{1200}{K} in \SI{25}{K} steps. At each temperature, a linear program finds the
lowest-free-energy mixture of competing solids, excluding the candidate, that
conserves all elemental fractions. Temperatures are reported only where the
candidate fit is usable and all controlling phases are available, and zero
crossings are reported as sampled brackets.

\subsection*{Data availability}
Training labels are drawn from public datasets: Matbench/Materials Project
\cite{matbench,materialsproject}, Alexandria-SCAN \cite{alexandriascan},
Alexandria electron--phonon \cite{alexandriaeph}, and SNUMAT \cite{snumat}.
Generated structures, first-principles inputs and outputs, and
finite-temperature free-energy data will be made publicly available when the
manuscript is published.

\subsection*{Code availability}
The DEFEAT implementation is built on FairChem \cite{fairchem}, modified to
support effective-atom inputs to EquiformerV3 \cite{equiformerv3}. Code and
campaign configurations will be made publicly available when the manuscript is
published.

\subsection*{Acknowledgements}
C.M.\ and T.A.A.\ acknowledge support from the U.S. National
Science Foundation under Award PHY-1549132, the Center
for Bright Beams. This work was supported in part by the U.S. National Science
Foundation through the Quantum Leap Challenge Institute for Manufacturable and
Resilient superconducting QUantum Information Systems (MARQUIS), under Award
No. OSI-2553633. J.T.\ acknowledges support from the
Webb@KIC Graduate Fellowship at the Kavli Institute
at Cornell for Nanoscale Science.

\subsection*{Author contributions}
C.M.\ and J.T.\ developed the method; C.M.\ performed the calculations; T.A.A.\ supervised
the work. C.M.\ wrote the manuscript with input from all authors.

\subsection*{Competing interests}
The authors declare no competing interests.

\bibliographystyle{unsrt}
\bibliography{refs}
\end{document}